\documentclass[twocolumn,11pt]{aastex631}

\usepackage{graphicx,amsmath}
\usepackage{color}
\usepackage{booktabs}
\usepackage{dutchcal}
\usepackage{multirow}
\usepackage{amsmath}
\usepackage{cuted}

\newcommand{\adsurl}[1]{\href{#1}{ADS}}
\providecommand{\url}[1]{\href{#1}{#1}}

\newcommand{\be}{\begin{equation}}
\newcommand{\ee}{\end{equation}}
\newcommand{\bea}{\begin{eqnarray}}
\newcommand{\eea}{\end{eqnarray}}

\usepackage{color}
\usepackage{ifthen}
\newboolean{editorial}
\setboolean{editorial}{true}
\newcommand{\editorial}[2]{\ifthenelse{\boolean{editorial}}{\textcolor{red}{[\textsf{\textbf{{#1}}}: }\textcolor{blue}{\textsf{{#2}}}\textcolor{red}{]}}{}}

\shorttitle{Lensed TDEs}
\shortauthors{Chen et al.}

\begin{document}

\title{Detectability of Strongly Gravitationally Lensed Tidal Disruption Events}

\correspondingauthor{Youjun Lu}
\email{luyj@nao.cas.cn}

\author[0000-0001-7952-7945]{Zhiwei Chen}
\affiliation{National Astronomical Observatories, Chinese Academy of Sciences, 20A Datun Road, Beijing 100101, China; 
}
\affiliation{School of Astronomy and Space Sciences, University of Chinese Academy of Sciences, 19A Yuquan Road, Beijing 100049, China}
\author[0000-0002-1310-4664]{Youjun Lu
}
\affiliation{National Astronomical Observatories, Chinese Academy of Sciences, 20A Datun Road, Beijing 100101, China; 
}
\affiliation{School of Astronomy and Space Sciences, University of Chinese Academy of Sciences, 19A Yuquan Road, Beijing 100049, China}
\author{Yunfeng Chen}
\affiliation{School of Astronomy and Space Sciences, University of Chinese Academy of Sciences, 19A Yuquan Road, Beijing 100049, China}
\affiliation{National Astronomical Observatories, Chinese Academy of Sciences, 20A Datun Road, Beijing 100101, China; 
}

\begin{abstract}
More than one hundred tidal disruption events (TDEs) have been detected at multi-bands, which can be viewed as extreme laboratories to investigate the accretion physics and gravity in the immediate vicinity of massive black holes (MBHs). Future transient surveys are expected to detect several tens of thousands {of} TDEs, among which a small fraction may be strongly gravitationally lensed by intervening galaxies. In this paper, we statistically etsimate the detection rate of lensed TDEs, with dependence on the limiting magnitude of the transient all-sky surveys searching for them. We find that the requisite limiting magnitude for an all-sky transient survey to observe at least $1$\,yr$^{-1}$ is $\gtrsim 21.3$, $21.2$, and $21.5$\,mag in the u-, g-, and z-bands, respectively. If the limiting magnitude of the all-sky survey can reach $\sim 25-26$\,mag in the u-, g-, and z-bands, the detection rate can be upto about several tens to hundreds per year. The discovery and identification of the first image of the lensed TDE can be taken as an early-warning of the second and other subsequent images, which may enable detailed monitoring of the pre-peak photometry and spectroscopy evolution of the TDE. The additional early-stage information may help to constrain the dynamical and radiation processes involving in the TDEs.

\end{abstract}


\section{Introduction}
\label{sec:intro}

A tidal disruption event (TDE) is a phenomenon caused by that a star moving too close to an massive black hole (MBH) is torned apart and then accreted by the MBH \citep{1975Natur.254..295H, 1988Natur.333..523R}, with tremendous radiation in multibands, including the optical/UV, X-ray, and radio bands \citep[e.g.,][]{2015JHEAp...7..148K, 2020SSRv..216..124V, 2021ARA&A..59...21G}. It is expected that the event rate of TDEs is $\sim 10^{-6}-10^{-3}$ yr$^{-1}$ per galaxy and there are many TDEs occured at each time in the universe \citep[e.g.,][]{1999MNRAS.309..447M, 2004ApJ...600..149W, 2006ApJ...645.1152H, 2013ApJ...774...87V, 2016MNRAS.455..859S, CYL20tde}. The detection and monitoring of these TDEs are of great importance as they may provide important information for understanding the MBHs and their growth in quiescent galaxies \citep{2003MNRAS.339..189Y,2006ApJ...652..120M}, the non-steady evolution of disk accretion and jet formation, as well as measuring the spacetime of the MBH \citep{2016MNRAS.463.2242X, 2019ApJ...872..151M}. 

More than one hundred TDEs have been detected by different telescopes via the unique light curves (LCs) of TDEs in {the X-ray by the ROentgen SATellite (ROSAT) and extended ROentgen Survey with an Imaging Telescope Array on Spectrum-Roentgen-Gamma mission (SRG/eROSITA)}, the UV band by the Galaxy Evolution Explorer (GALEX), the optical band by the Solan Digital Sky Survey (SDSS), the Panoramic Survey Telescope and Rapid Response System (PAN-STARRs), and the Zwicky Transient Facility (ZTF), etc. \citep[e.g.,][]{1999A&A...349..389V,2002AJ....124.1810S,2005ApJ...619L...1M,2008AJ....135..338F,2008ApJ...678L..13K,2011MSAIS..17..159C,2016arXiv161205560C,2019PASP..131a8002B}. Future transient surveys, such as those by Vera C. Rubin telescope (hereafter Rubin) and SiTian are aimed to detect more than several tens thousands such TDEs \citep[][also \citealt{2021AnABC..93..628L}]{2020ApJ...890...73B}. It is possible that a small fraction of these TDEs may be strongly lensed by intervening galaxies with double or even quadruple images, similar as for the cases of Quasars, supernovae, gravitational wave sources, etc. \citep[e.g.,][]{2010MNRAS.405.2579O, 2018MNRAS.476.2220L}. We expect that these lensed TDEs are useful tools for understanding the underlying physical processes involved in the non-steady evolution of the TDE disk accretion and jet formation as their multiple images may provide more detailed LC information than that of normal TDEs. However, it is not yet clear how many such lensed TDEs can really be detected by future transient surveys and whether addition information can be obtained from the lensed TDE images for understanding the underlying physics.

In this paper, we investigate the lensing of TDEs and estimate the detection rate of lensed TDEs by assuming the TDE survey strategy. The paper is organized as follows. In Section~\ref{sec:method}, we briefly introduce the method to estimate the detection rate of lensed TDEs, by considering the cosmic distribution of TDEs (Section~\ref{sec:tde}), the modelling of TDE LCs (Section~\ref{sec:lc}), the lensing statistics, and observation strategies (Section~\ref{sec:lens}). Our main results are presented in Section~\ref{sec:results}. Discussions and conclusions are given in Section~ \ref{sec:conclusion}.

Throughout the paper, we adopt the cosmological parameters as $(h_0,\Omega_{\rm m},\Omega_\Lambda)=(0.68,0.31,0.69)$ \citep{Aghanim2020}.

\section{method}
\label{sec:method}

In this section, we introduce the methods to estimate the cosmic evolution of TDE rate, the lensing rate of TDEs, the TDE light curves, and the rate of lensed TDEs and their detection as detailed below. 

\subsection{TDE distributions}
\label{sec:tde}

We generate the TDE samples based on \citet{CYL20tde}, in which the volumetric rate of cosmic TDEs are obtained by detailed modelling of the loss-cone and loss region filling due to both the two-body relaxation and the stellar orbital precession in nonspherical potentials [see Eq.~(17) and Fig.~6 therein] \citep[see also][]{2020ApJ...897...86C, 2002MNRAS.331..935Y}. The TDE phenonmena is suppressed when the mass of the central MBH is large ($M_\bullet \gtrsim 10^8 M_\odot$ if the MBH is nonspinning or $\gtrsim 10^9 M_\odot$ if it is maximumly rotating) because the stars that can approach the immediate vicinity of the MBH cannot be disrupted before they are directly swallowed (e.g., \citealt{Kesden12, Coughlin22}). In this paper, we assume that all MBHs are the non-rotating Schwarzschild MBHs, for simplicity. We estimate the volumetric TDE rate according to \citet{CYL20tde} by considering the observable TDE fraction given by \citet{Coughlin22}. The observable fraction was obtained by considering only the disrupted stars via the full loss-cone regime. 

Note that the volumetric stellar TDE rate is estimated to be $\sim 3\times 10^{-5}\rm yr^{-1} Mpc^{-3}$ for MBHs with $M_{\bullet} \sim 10^{5}-10^{8} M_{\odot}$ at $z=0$ in \citet[][]{CYL20tde}, which is higher than the observational constraint ($\sim 10^{-6} \rm yr^{-1} Mpc^{-3}$) inferred from TDE candidates with MBH mass measurements \citep{2020SSRv..216..124V,2023ApJ...955L...6Y}. The difference between the theoretical estimate and the observational constraint may be explained by the incompleteness of the observed TDE mass distribution {(about $33$ events) or the un-determined MBH occupation fraction in low-mass galaxies. For the TDE observations \citep[e.g.,][]{2023ApJ...955L...6Y}, currently only three reported TDEs have central MBHs with masses $<10^{5.5}M_\odot$, i.e.,  AT2020neh \citep{2022NatAs...6.1452A}, AT2020wey \citep{2020TNSCR1080....1A} and AT2021yte \citep{2021TNSCR3611....1Y} , and the cases with smaller masses (and even a fraction of the cases with masses $>10^{5.5}M_\odot$) may be missed due to selection effects.} For the theoretical estimation of TDE rates, it was assumed that the mass range can extend to $10^5M_\odot$ and the occupation fraction of MBH in small galaxies is $100\%$. Therefore, the adopted TDE rates, and consequently our estiamtes on the detectable lensed TDE rates, can be taken as an optimitic estimation, which could be overestimated by a factor of a few but at most an order of magnitude. 

\begin{figure}
\centering
\includegraphics[width=1.0\columnwidth]{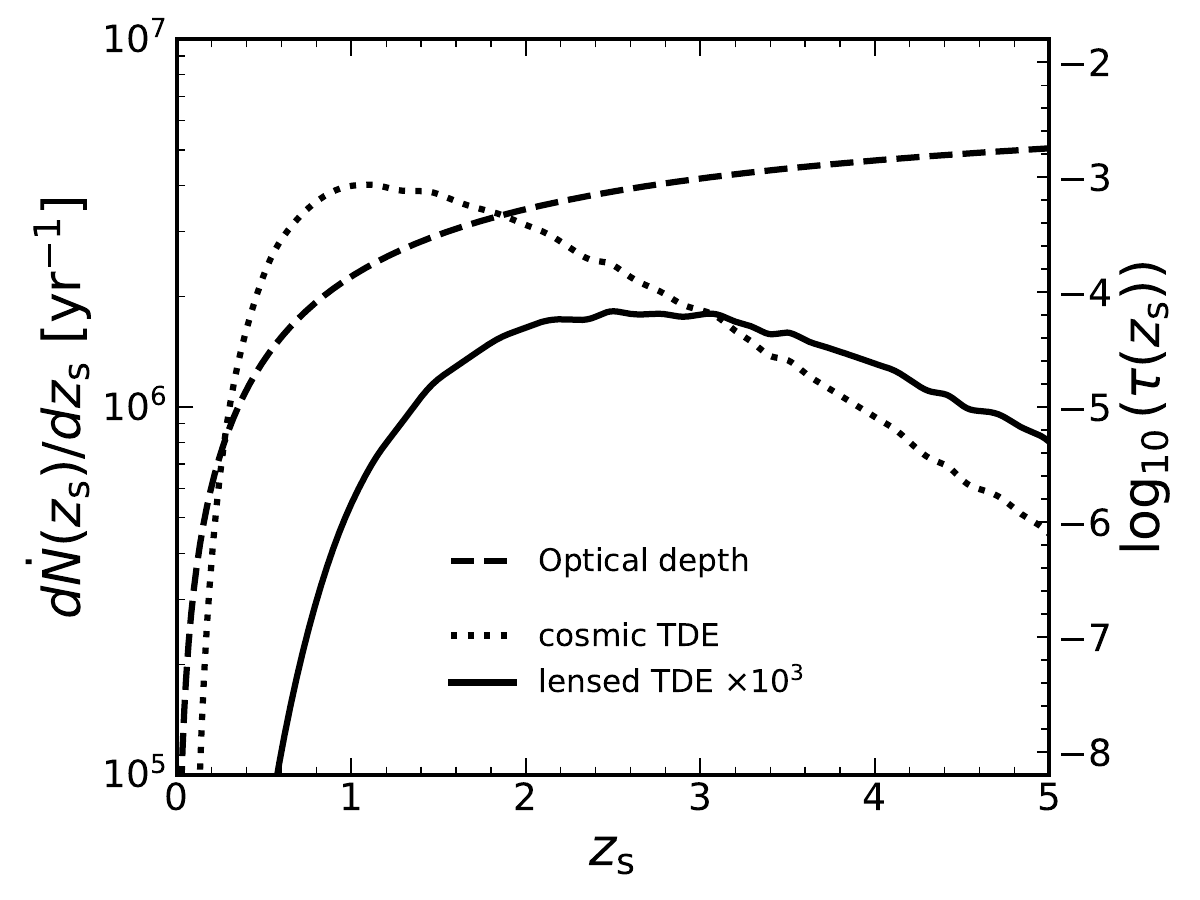}
\caption{
The evolution of the cosmic TDE rate and the optical depth of gravitational lensing. The dotted line shows the number density per unit redshift $\frac{d\dot{N}(z_{\rm s})}{dz_{\rm s}}$ of the cosmic TDE, and the dashed line represents the optical depth evolution $\tau(z_{\rm s})$ obtained from the SIE+external shear model. The solid line shows the number density of lensed TDEs per unit redshift times a factor of $10^3$. 
}
\label{fig:tau}
\end{figure}

Figure~\ref{fig:tau} shows the number density per unit redshift $\frac{d\dot{N}(z_{\rm s})}{dz_{\rm s}}$ evolution of the cosmic TDE with redshift. As seen from this figure, $\frac{d\dot{N}(z_{\rm s})}{dz_{\rm s}}$ peaks at $z_{\rm s}\sim 0.8$ and declines rapidly with increasing redshift. According to this number density distribution, there are totally $\sim 10^7$ TDEs occured within a year in the whole universe. One can generate mock samples for such TDEs occurred at different times ($z_{\rm s}$) in galaxies with different central MBHs ($M_\bullet$) according to \citet{CYL20tde} by the Monte Carlo method.

As for the disrupted stars of the mock TDE samples, we adopt the polytropic stellar model with index $n=3$ $(\gamma=4/3)$ for typical main sequence stars \citep{1994sse..book.....K}. In this case, the radial density profile of the star $\rho(R)$ can be written in terms of $m_\star$ and $x_\star$ (i.e., the mass and radius of a star in units of solar mass and solar radius) as: 
\begin{equation}
\rho(R)=\rho_{\rm c}w^3(z),
\end{equation}
where $z=6.896R/x_{\star}$ is a dimesionless variable, $\rho_{\rm c}=54.18 m_{\star}/(4\pi x_{\star}^3/3)$ is the central density, and $w$ is the solution of the Lane-Emden equation 
\begin{equation}
\frac{d^2w}{dz^2}+\frac{2}{z}\frac{dw}{dz}+w^3=0,
\end{equation}
with intial condition $w(0)=1$ and $w^{\prime}(0)=0$.

The mass distribution of the disrupted stars are assumed to be propotional to the initial star mass function proposed by \citet{2001MNRAS.322..231K}.\footnote{Note here that we only consider the abundance difference of different type of stars for simplicity, but ignore the dependence of the loss-cone draining/refilling rate on the properties of the disrupted stars, e.g., mass and different evolution stages, for simplicity. One may find some discussions on such dependence in \citet{2023ApJ...952..135C}.} We apply the mass-radius relation $x_\star=m_\star^\delta$ to estimate the radius of 
\textbf{stars}  with $\delta=0.8$ if $m_\star\leq 1$ and $\delta=0.5$ if $m_\star>1$ \citep[see][]{1994sse..book.....K}. Note that the above simplified approximation of mass-radius relationship is only valid for the main-sequence stars.

By adopting the Gibbs Monte-Carlo sampling method \citep[e.g., ][]{gibbs}, we randomly generate $10^6$ TDE events with different parameters, i.e., $(M_\bullet, z_{\rm s}, m_{\star}, x_{\star})$ for latter analysis. Here we limit the redshift range of the mock TDEs to $z \in (0, 5)$ and their MBH mass $M_\bullet$ larger than $10^5 M_\odot$. Note that the cosmic TDE rate drops quickly at the high mass end $ M_{\bullet}\gtrsim 10^7 M_{\odot}$, which results in $M_{\bullet}$ in the our samples to be approximately limited within the range of $(10^5 M_{\odot},10^8 M_{\odot})$. 

\subsection{TDE Light curves}
\label{sec:lc}

Accurately predicting the light curves (LCs) of a TDE remains challenging due to the complicated physical processes involved in it, including the circularization of the fallback debris, the evolution of the accretion disk, the disk wind, and the radiative transfer, etc. For demonstration purpose, in this paper we adopt simple semi-analytical models to estimate TDE LCs, of which the observable LCs first rapidly increase due to the starting of the disk accretion, then reach a peak mainly caused by the super-Eddington wind, and decline later due to the time-dependent accretion of the disk. Below we only summarize the main equations for the predicting TDE LCs. For more details one can find in \citet{2009MNRAS.400.2070S}, \citet{2011MNRAS.410..359L} and \citet{2023PASP..135c4102K}. 

\begin{table*}
\caption{Model parameters adopted for generating the light curves of TDEs }
\label{table:para} 
\centering
\begin{tabular}{lccc} \hline    \hline
       $z_s$ & TDE redshift & $[0,5]$ & \citet{CYL20tde}\\
       $M_\bullet$ & MBH mass & $[10^5,10^{7.5}] \rm M_{\odot}$ &  \citet{CYL20tde}\\
       $m_*$ & star mass &$[0.08,10] \rm M_{\odot}$ &  \citet{2001MNRAS.322..231K}  \\
       $x_*$ & star radius &$[0.13,3.16] \rm R_{\odot}$& \citet{1994sse..book.....K} \\
       $\beta$ & penetration factor& $[0.8,\infty)$ & \citet{2022ApJ...933...96Z}   \\    \hline  \hline
       $\eta$ & radiative efficiency & 0.1& \citet{2023PASP..135c4102K}  \\
       $f_{\mathrm{v}}$ & wind velocity constant & 1.0 & \citet{2023PASP..135c4102K}  \\
       $f_{\text {out }}$ &  energy fraction of wind & [0.1,0.7) & \citet{2011MNRAS.410..359L}\\ \hline    \hline
\end{tabular}
\tablecomments{
The first, second, and third columns list the name, physical meaning, and the range of the model parameters. More detailed information on each parameter can be seen in the corresponding literature listed in the fourth colunm.}
\end{table*}

In this simplified scenario, we assume that the cirularization timescale for fallback material is shorter than the fall-back timescale. Then the time of the first debris particle returning to the pericente $t_{\rm min}$ is roughly the elapsed time between the moment of disruption and the start of the mass accretion onto the central MBH, approximately the period of this particle that is the closest to the MBH when the star enters the periapsis distance ($r_{\rm p}$), i.e.,
\begin{equation}
t_{\rm min}=\frac{\pi}{\sqrt{2}}\left(\frac{r_{\rm p}}{R_{\star}}\right)^{3/2}\sqrt{\frac{r^3_{\rm p}}{GM_\bullet}},
\label{eq:class}
\end{equation}
where $R_\star$ is the radius of the star. We can define a penetration parameter $\beta= r_{\rm t}/r_{\rm p}$ with $r_{\rm t}=\left(M_\bullet/m_{\star}\right)^{1/3} R_{\star}$ representing the tidal radius of the star. This penetration parameter denotes the dynamical behaviour of the disrupted stars and it has significant effect on the output radiation of TDEs. Normally, different TDEs can have different $\beta$. In this paper, we adopt the probability distribution of $\beta$ obtained from numerical simulations by \citet{2022ApJ...933...96Z}, i.e., $P_\beta \propto \beta^{-2}$ for full TDEs with $\beta>0.8$.\footnote{If $\beta$ is slightly larger than $0.8$, they are partial TDEs.} Note that the accurate split value of $\beta$ between the full and partial TDE regime is still in debate due to the complexities in the disruption processes. The definition of the minimum time $t_{\rm min}$ can be slightly different from the classical $r_{\rm t}$ description in the literature \citep[e.g.,][]{2009MNRAS.400.2070S, 2011MNRAS.410..359L}.

To avoid the ambiguity, \citet{2023PASP..135c4102K} modified the above equation for $t_{\rm min}$ as 
\begin{equation}
t_{\rm min}=\frac{\pi}{\sqrt{2}}\left(\frac{r'_{\rm t}}{R_{\star}}\right)^{3/2}\sqrt{\frac{{r'_{\rm t}}^3} {GM_\bullet}},
\label{eq:tmin}
\end{equation}
where $r'_{\rm t}=r_{\rm t}/\beta_{\rm d}=\beta r_{\rm p}/\beta_{\rm d}$ and $\beta_{\rm d}$ is a free constant parameter.
When $\beta_{\rm d}=\beta$, it restores the classical definition shown by Equation~\eqref{eq:class}. When $\beta_{\rm d}=1.20$ and $1.85$, the resulted LCs are consistent with the simulation ones obtained for the disruption of white dwarfs and main sequence stars, respectively \citep[see][]{2013ApJ...767...25G}. We shall see later that the peak luminosity $L_{\rm \nu,p}$ of a TDE is strongly dependent on the choice of $\beta_{\rm d}$. For comparison, we consider three different values of $\beta_{\rm d}$ in our calculations, i.e., $1.00$, $1.20$, and $1.85$, respectively.

Given the minimum time $t_{\rm min}$, one may estimate the time-dependent mass fallback accretion rate $\dot{M}_{\rm fb}$ of a disrupted star as \citep[][]{2018ApJ...857..109G}
\begin{equation}
\dot{M}_{\mathrm{fb}}=\frac{4 \pi}{3} \frac{x_\star}{t_{\min }}\left(\frac{t}{t_{\min }}\right)^{-5 / 3} \int_{x(t)}^{x_\star} \rho(R) R d R
\end{equation}
where $x(t)=x_{\star}(t/t_{\rm min})^{-5/3}$ and $\rho(R)$ is the density profile of the star. The multiband LCs of a TDE event can be then estimated by this fallback accretion rate $\dot{M}_{\rm fb}$. 

As for the super-Eddington wind, we assume that it is launched at $r_{\rm L}$ with a velocity of
\begin{equation}
v_{\rm w}=f_v \left( \frac{2GM_\bullet}{r_{L}}\right)^{1/2},
\end{equation}
where $f_v$ is a constant parameter. The photosphere radius of the wind, $r_{\rm ph}$,  can be calculated as:
\begin{equation}
r_{\rm ph}=\frac{\kappa f_{\rm out} \dot{M}_{\rm fb}}{4\pi v_{\rm w}},
\end{equation}
where $\kappa$ is the mean opacity and $\kappa \sim \kappa_{\rm es}$ \citep{2021MNRAS.502.3385M} in this case, $f_{\rm out}$ is the fraction of the fallback material that is ejected away as outflow with velocity $v_{\rm w}$.

Assuming the outflow is adiabatic, then the temperature of the photosphere is
\begin{equation}
T_{\mathrm{ph}}=T_{\mathrm{L}} \left(r_{\mathrm{ph}} / r_{\mathrm{L}}\right)^{-2 / 3} \left(f_{\text {out }} / f_v\right)^{1 / 3},
\end{equation}
where $T_{\mathrm{L}}$ is the temperature of the photosphere at $r_{\rm L}$. Therefore, the luminosity of the super-Eddington wind can be estimated as
\begin{equation}
L_{\mathrm{w}}(\nu)=4 \pi^2 r_{\mathrm{ph}}^2 B_\nu\left(T_{\mathrm{ph}}\right),
\end{equation}
assuming that the photosphere is optically thick and its radiation spectrum is roughly a blackbody described by $B_\nu(T_{\rm ph})$.

On the other hand, following the steady slim disk model proposed by \citet{2023PASP..135c4102K}, we assume that 1) the fallback timescale is substantially longer than the viscous timescale; 2) the radiation pressure is much larger than the gas pressure; 3) the viscous stress is proportional to the radiation pressure. Then, the temperature of the disk should vary with the radius $r$ as \citep{2009MNRAS.400.2070S}:
\begin{eqnarray}
 & T(r)  & = \left( \frac{3 G M_\bullet \dot{M}_\bullet f}{8 \pi r^3 \sigma_{\mathrm{SB}}} \right)^{1/4} \times \nonumber \\
 & & \left[\frac{1}{2}+\left\{\frac{1}{4}+\frac{3}{2}  f \left(\frac{\dot{M}_\bullet}{\eta \dot{M}_{\mathrm{Edd}}}\right)^2\left(\frac{r}{R_{\mathrm{S}}}\right)^{-2}\right\}^{1 / 2}\right]^{-1/4},
\end{eqnarray}
where $\dot{M}_\bullet=(1-f_{\rm out})\dot{M_{\rm fb}}$ is the disk accretion rate, $f=1-\sqrt{R_{\rm in}/r}$, $\eta$ is the radiative efficiency.

Here we adopt the inner edge of disk to be the last stable circular orbit, $R_{\rm in}=3R_{\rm s}$, by assuming $a=0$ for simplicity, and the outer edge to be the circularization radius, $R_{\rm out}=2r_{\rm p}$. The luminosity of the disk is 
\begin{equation}
L_{\mathrm{d}}(\nu)=2 \int_{R_{\text {in }}}^{R_{\text {out }}} 2 \pi^2 B_\nu(T(r)) r d r. 
\end{equation} 
Instead of simply summing up $L_{\rm d}$ and $L_{\rm w}$, we consider the reprocessing of the disk luminosity into the optical radiation by the wind particles as that discussed in \citet{2023PASP..135c4102K}. As the lensed TDEs are mostly at high redshift, we also consider the K-correction for LC observations at each band by using the predicted TDE spectra $L_{\rm d}(\nu)$. We summarize here that the parameters in our model for TDE events include the TDE redshift $z_{\rm s}$, the MBH mass $M_\bullet$, the mass and radius of the disrupted star ($m_\ast$ and $x_\ast$), the penetration factor $\beta$, the radiative efficiency $\eta$, the strength of the wind velocity $f_v$, and the energy fraction of the super-Eddington wind $f_{\rm out}$. {Here we also note that the LCs of a TDE depend weakly on $\eta$ and $f_v$, but strongly on $f_{\rm out}$. In this paper, we adopt the fitting formalism of $f_{\rm out}$ given by \citet[][see Eq. 28 therein]{2011MNRAS.410..359L}), which only considers the dependent on the mass fall-back rate but not $f_{\rm out}$. Therefore, we argue that the predicted LCs are not sensitive to these three model parameters, but in turn directly related to the intrinsic parameters of the MBH and disrupted stars. For clarity, we list the setting for these parameters in Table~\ref{table:para}.}

\begin{figure}
\centering
\includegraphics[width=1.0\columnwidth]{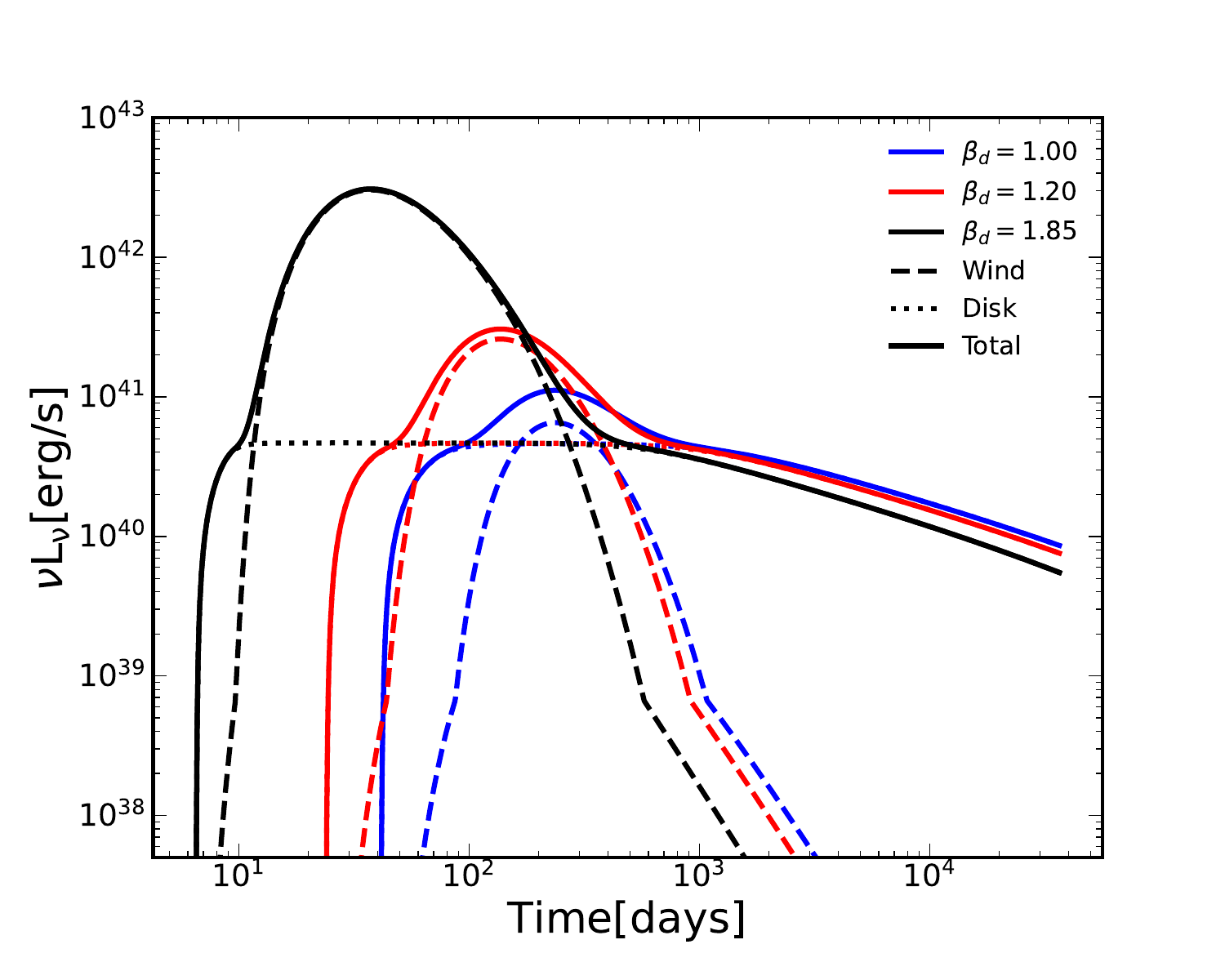}
\caption{The estimated intrinsic light curves of TDEs obtained from the Wind+Disk model, by assuming $\beta=1$, $\eta=0.1$, $f_{\rm out}=1$, $f_{v}=1$, $M_\bullet=10^6 M_\odot$, and the mass and radius of the disrupted star $m_{\star}=1\rm M_\odot$ and $r_{\star}=1 R_\odot$, respectively. The dashed and dotted lines show the radiation components of the disk and the wind in the g-band, respectively, while the solid lines represent the after-processing total radiation in the g-band. Different colors represent different choices of the free parameter $\beta_{\rm d}$, i.e., blue for $\beta_{\rm d}=1.00$, red for $\beta_{\rm d}=1.20$, and black for $\beta_{\rm d}=1.85$.}
\label{fig:LC}
\end{figure}

We show the g-band LCs estimated from the above simple model for an example TDE in Figure~\ref{fig:LC} for different choices of the $\beta_{\rm d}$ value. The dashed and dotted lines in this figure show the radiation contributed respectively by the disk and the wind components of the TDE, and the solid line shows the after-processing total radiation. Blue, red, and black lines are for the cases with $\beta_{\rm d}=1$, $1.2$, and $1.85$, respectively. As expected, the wind radiation dominates at the early stage of the TDE and the disk radiation dominates the late stage. Different choices of $\beta_{\rm d}$ may not affect the radiation of disk but have a strong effect on the radiation of the wind. The set of a larger $\beta_{\rm d}$ lead to a higher peak luminosity $L_{\rm \nu,p}$. The peak luminosity for a TDE with $\beta_{\rm d}=1.85$ is larger that with $\beta_{\rm d}=1$ by a factor of several tens. In addition, the timescale for the TDE to reach its peak luminosity at the early stage decreases by a factor of several tens when $\beta_{\rm d}$ increases from $1$ to $1.85$. Therefore, the detection rate of lensed TDEs are strongly dependent on the choice of $\beta_{\rm d}$. In this paper, we chose two different values for $\beta_{\rm d}$, i.e., $1.0$ and $1.85$, to focus on the disruption of main sequence stars, but ignore the case with $\beta_{\rm d}=1.20$ for white dwarfs. Here we note that when $\beta_{\rm d}=1.0$, Equation~\eqref{eq:tmin} restores the typical $t_{\rm min}$ defined by the tidal radius $r_{\rm t}$ in \citet{2013MNRAS.435.1809S}, and the case for $\beta_{\rm d}=1.85$ is retained based on the simulation of \citet{2013ApJ...767...25G}.

\begin{figure}
\centering
\includegraphics[width=1.0\columnwidth]{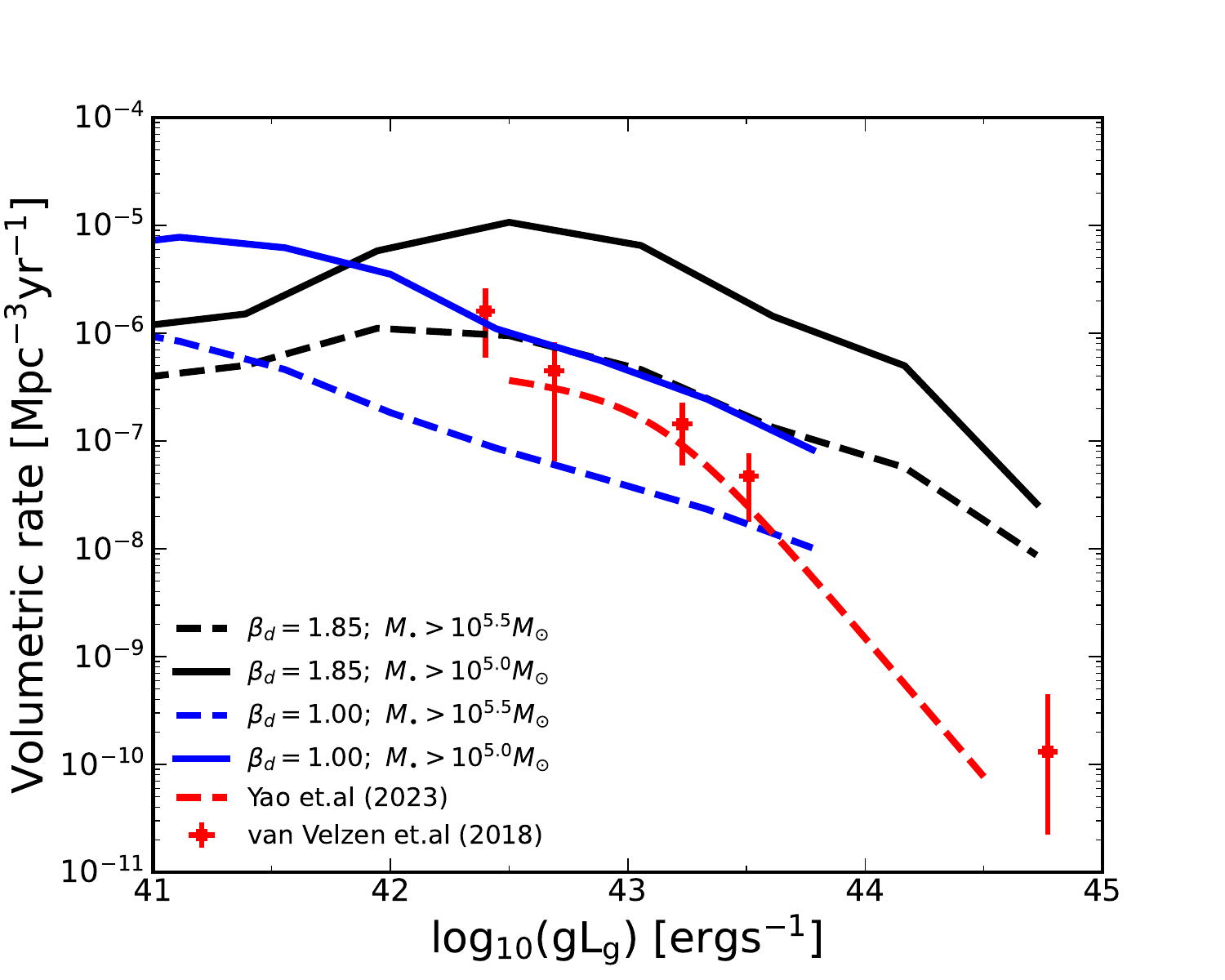}
\caption{The luminosity function of the mock TDE population within $z_{\rm s}<0.2$ generalized as discussed in section \ref{sec:tde} and \ref{sec:lc}, which is generally consistent with current observation.  The black solid and dashed lines represent the results for TDEs with $M_{\bullet}>10^{5.0}M_{\odot}$ and $M_{\bullet}>10^{5.5}M_{\odot}$ when $\beta_{d}=1.85$, respectively. The blue lines are similar, except for $\beta_{d}=1.00$. For comparison, We also plot the observational constraint of TDE luminosity function from \citet{2018ApJ...852...72V} and the double power-law fitted results in  \citet{2023ApJ...955L...6Y}.}
\label{fig:LF}
\end{figure}

To validate the theoretical LC model adopted in this paper, as shown in Figure~\ref{fig:LF}, we further compare the TDE luminosity function obtained from our model for the mock nearby TDEs (i.e., $z_{\rm s}<0.2$) with the observational ones \citep[e.g., ][]{2018ApJ...852...72V,2023ApJ...955L...6Y}. As seen from this figure, the TDE lumonisity function obtained from our model is roughly consistent with the observations, if limiting the mock samples with a similar MBH mass range as the observational ones, i.e., $M_\bullet>10^{5.5}M_\odot$. This consistence suggests that the TDE population model and the LC model we adopted are reasonable and supports the application of these models to estimate the detection rate of strongly gravitationally lensed TDEs adequately. 

\subsection{Lensing statistics}
\label{sec:lens}

To estimate the detection rate of strongly gravitationally lensed TDEs,
the singular isothermal ellipsoid profile (SIE) with external shear is adopted for the lens model as the galaxy-galaxy strong lensing is mainly due to elliptical galaxies \citep[e.g.,][]{1984ApJ...284....1T, 2007MNRAS.379.1195M}. In this model, the double image cases are the most common circumstance \citep[e.g.,][]{2010MNRAS.405.2579O, 2018MNRAS.480.3842O}. The critical quantity for estimating the lensed rate is the lensing optical depth $\tau(z_{\rm s})$ given by \citep[e.g.,][]{2018MNRAS.480.3842O, 2018MNRAS.476.2220L, 2023ApJ...953...36C}
\begin{equation}
\begin{aligned}
\tau(z_s)=&\frac{1}{4 \pi} \int_{0}^{z_s} \frac{dV({z_{\rm l}})}{dz_{\rm l}} dz_l \iint d\gamma_{1}d\gamma_{2}P_{\boldsymbol{\gamma}}(\gamma_{1},\gamma_{2}) \int de P_{\rm e}(e) \\
& \times  \int d\sigma_{\rm v} \frac{dn(\sigma_{\rm v},z_{\rm l})}{d\sigma_{\rm v}}  S_{\rm cr}(\sigma_{v},z_l,z_s,\gamma_{1},\gamma_{2},e), 
\end{aligned}
\label{eq:tau}
\end{equation}
where $z_{l}$ is the redshift of lens, and $P_{\rm e}$ is the axis-ratio distribution of the lens systems, which is assumed to be a truncated Gaussian distribution in the range of $[0.2,1]$ with a mean of $0.7$ and a standard deviation of $0.16$ to match with the observations on early-type galaxies \citep{2003ApJ...594..225S}.
And $P_{\boldsymbol{\gamma}}$ is the probability distribution of 
the two dimensional external shear in Cartesian coordinates, which represent the morphology and external environment near the line of sight of the lens. Following \citet{2005ApJ...624...34H}, we assume that the amplitude $\boldsymbol{\gamma}$ follows a log-normal distribution with a mean of $\ln{ 0.05}$ and a standard deviation of $0.2$, and the direction of $\boldsymbol{\gamma}$ is randomly distributed.


\begin{figure}
\centering
\includegraphics[width=1.05\columnwidth]{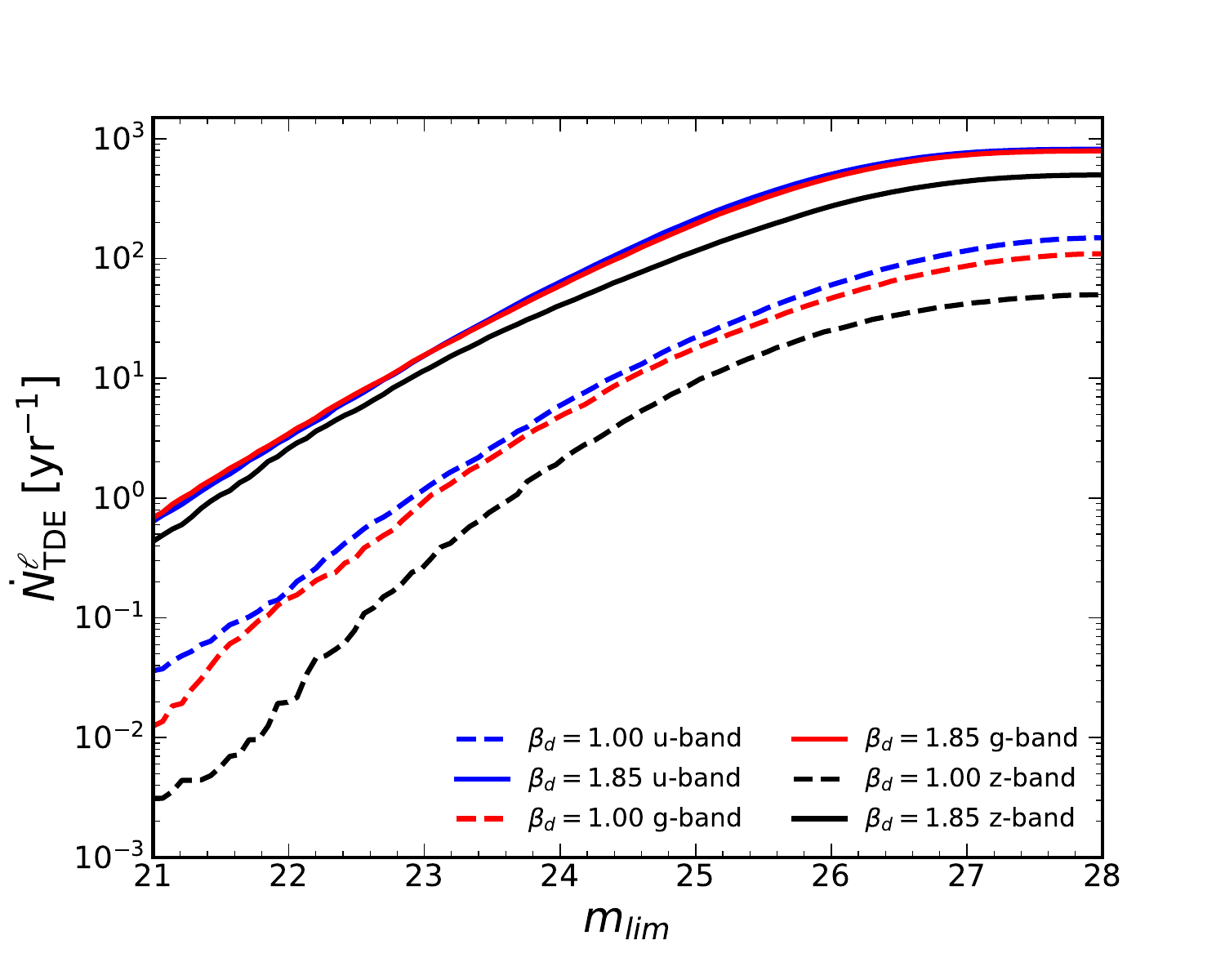}
\caption{
The detection rate of lensed TDEs $\dot{N}^{\mathcal{l}}_{\rm TDE}$ by surveys with various limiting magnitude $m_{\rm lim}$ in different bands. The blue, red, and black lines show the results  observed in u-, g-, and z-band, respectively, and the dashed and solid lines represent the LCs obtained by set $\beta_{\rm d}=1.00$ and $\beta_{\rm d}=1.85$, respectively.
}
\label{fig:rate}
\end{figure}

In Equation~\eqref{eq:tau}, the lense cross-section $S_{\rm cr}$ of the intervening galaxy is dependent on the velocity dispersion, the redshifts of lens and source, eccentricity and the external shear of nearby enviroment. For simplicity, we adopt the observational constraint on 
the  velocity distribution function (VDF) from \cite{2007ApJ...658..884C} and \cite{Piorkowska:2013eww}: 
\begin{equation}
\frac{dn(\sigma_{\rm v},z_{l})}{d\ln \sigma_{\rm v}}=n_z \frac{\beta}{\Gamma(\alpha/\beta)}
\left(\frac{\sigma_{\rm v}}{\sigma_z}\right)^{\alpha}\exp{\left[-\left(\frac{\sigma_{\rm v}}{\sigma_z}\right)^{\beta}\right]},
\label{eq:lens}
\end{equation}
and
\begin{equation}
{n_z = n_{0}(1+z)^{\kappa_{n}};\quad \sigma_z = \sigma_{\rm v0}(1+z)^{\kappa_{\rm v}}}.
\end{equation}
Here $\sigma_{\rm v0}$ is the characteristic velocity dispersion, $\alpha$ is the power-law index at the low-velocity end, $\beta$ is the high-velocity exponential truncated index, and $\Gamma(\alpha/\beta)$ is the Gamma function. The reference parameters in this formula are chosen to be the canonical values, i.e.,  $(n_0, \sigma_{\rm v0}, \alpha, \beta) = (0.008h^{3} {\rm Mpc}^{-3}, 161{\rm km \, s^{-1}}, 2.32, 2.67)$. As for the evolution parameters $\kappa_n$ and $\kappa_{\rm v}$, we adopt the fitting results from \citet{2021MNRAS.503.1319G}, i.e., $\kappa_n = -1.18$ and $\kappa_{\rm v}=0.18$.
The dashed line in Figure~\ref{fig:tau} shows the optical depth $\tau(z_{\rm s})$ evolution with redshift $z_{\rm s}$ calculated by Equation~\eqref{eq:tau} with the above approximations. As shown in Figure~\ref{fig:tau}, the higher the redshift of the source is, the larger the probability to be lensed will be. Moreover, from the dotted line, it is obvious that the intrinsic lensed TDEs number density per unit redshift peaks  at $z_{\rm s}\sim 2.5$ without any observational constraint.  

\begin{figure}
\centering
\includegraphics[width=1.0\columnwidth]{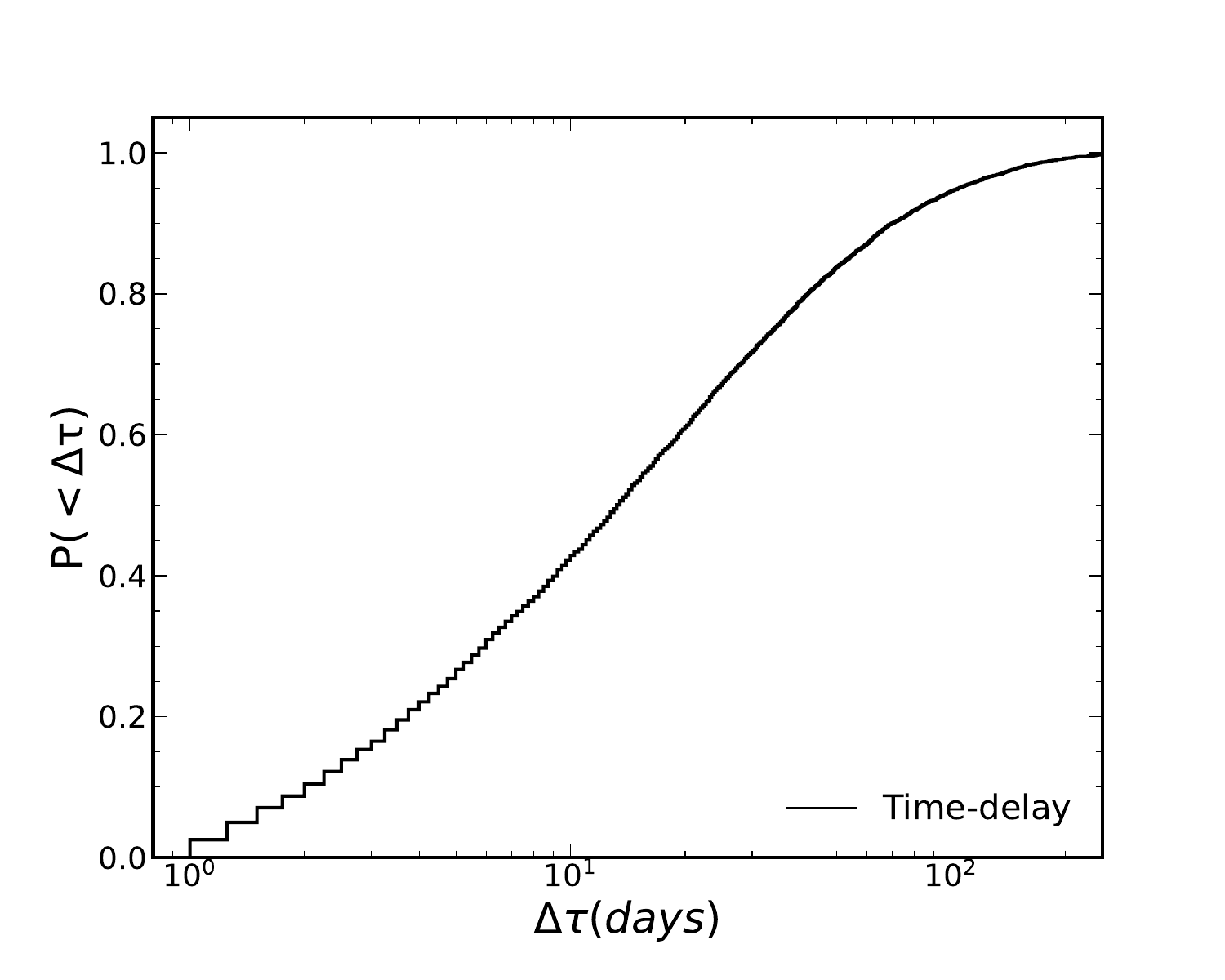}
\caption{
The cumulative probability distribution of the time-delay $\Delta \tau$ between the two images of lensed TDEs.  
}
\label{fig:td}
\end{figure}

\begin{figure*}
\centering
\includegraphics[width=0.45\textwidth]{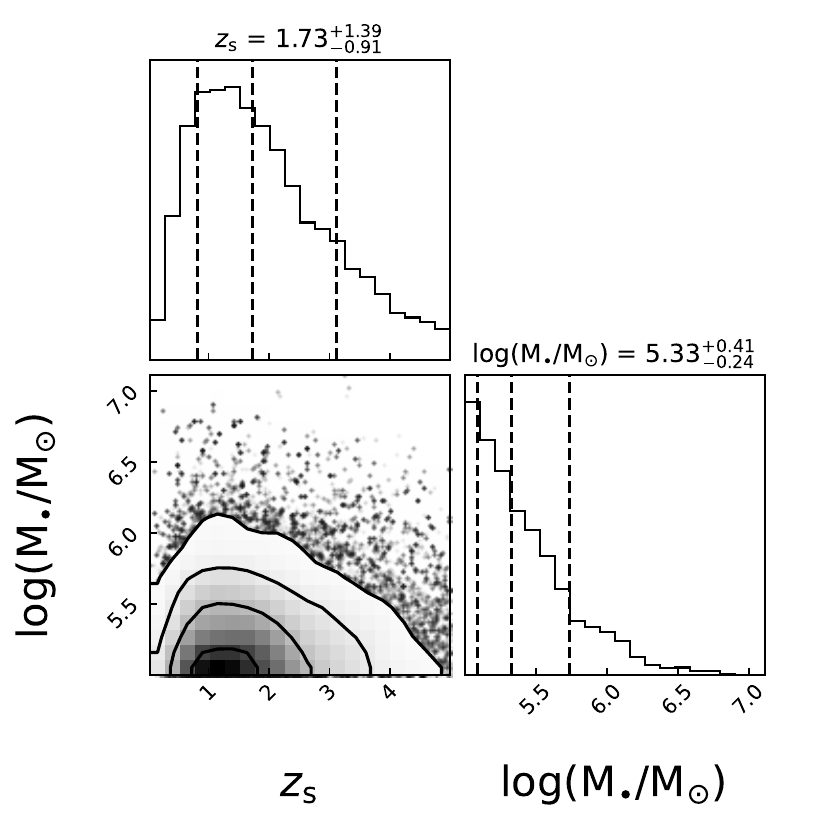}
\includegraphics[width=0.45\textwidth]{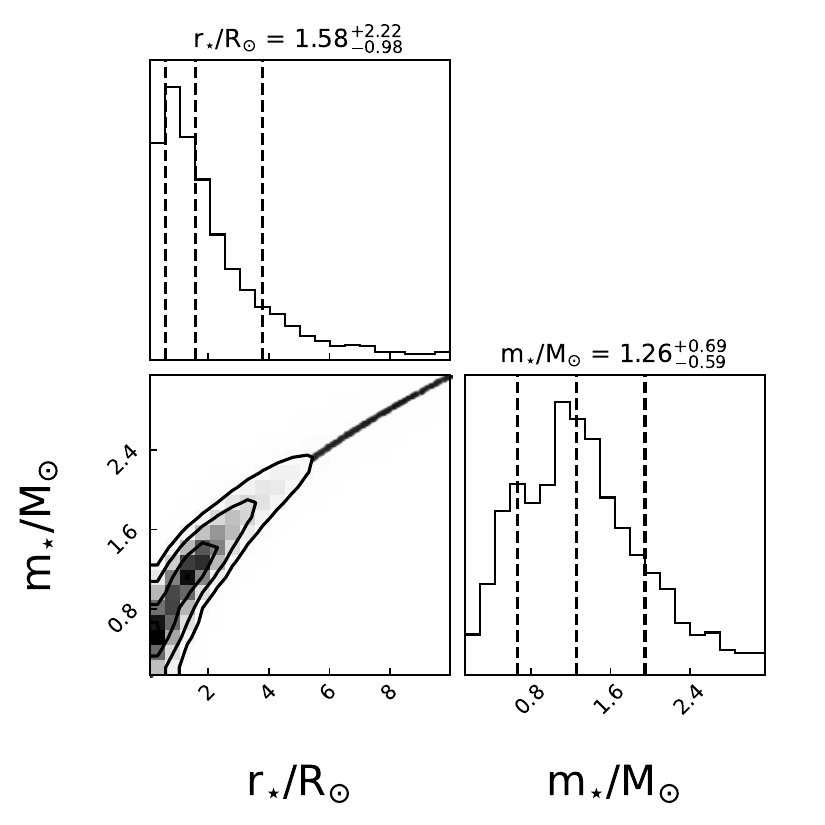}
\caption{
Distributions of the properties of the MBHs and disrupted stars in the lensed TDE samples detectable for an all-sky survey with limiting magnitude of $23.5$ mag in the g-band. Left and right panels show the distribution of the MBH mass $M_\bullet$ and redshift $z_{\rm s}$, and the distribution of the disrupted star radius $R_\star$ and mass $m_\star$, respectively. 
}
\label{fig:para}
\end{figure*}

There are two different cases in identifying such lensed TDEs considering of whether their lensed images can be spatially resolved. If the angular resolution of a survey searching for a TDE is sufficiently high so that its two images can be distinguished spatially from each other. (Here we ignore those lensed TDEs with quadruple images, as they are much rarer than those with double images \citep{2018MNRAS.476.2220L}.) On the other hand, if the angular resolution is not sufficiently high, the two images may be mixed together and cannot be spatially separated from each other, which may lead to a double peaked LCs due to the time-delay effect by the strong lensing. Here we define a lensed TDE as {\it Resolvable} if the angular separation of the two images $\Delta \theta$ is larger than $0.1''$, which is chosen as the designed angular resolution of J-band in RST. In our samples, more than $\sim 99\%$ of our lensed TDE samples are resolvable.
If chosen another band, such as i-band of CSST ($0.18''$) and VIS-band of Euclid ($0.23''$), the fraction will  decrease slightly but still larger than $\gtrsim 95\%$.

Before the presentation of our results, we propose the strategy for searching and detection of the lensed TDEs. First, we search for TDEs in a transient survey and compared their host galaxies with the galaxy-galaxy lensing banks produced by space sky-survey \citep{2015ApJ...811...20C, 2022ApJ...940...17C}, such as Chinese Space Station Telescope (CSST) and Euclid.
Here we optimisitically assume that all the lensed TDEs can have identifiable lensed host galaxies, i.e., $f_{\rm host}=1$, which will be discussed later in the discussion section. Therefore, for this reason, one can define a TDE detected in the transient survey in the center of a lensed host galaxy as the first image, and then apply telescopes with deeper limiting magnitudes to search for the corresponding second image. In this case, the limiting magnitude of the latter search of the LC peak can be higher, in principle, e.g., reach $\gtrsim 28$\,mag \citep[for example, James Webb Space Telescope (JWST) and Nancy Grace Roman Space Telescope (RST),][]{2006SSRv..123..485G,2013arXiv1305.5422S}. Once both the first and second images are detected, we count it as a `detectable' lensed TDE. Note that by this strategy, there is a time-sequential effect, i.e.,  that we have $50\%$ probability to find the second image and miss the first image in the first step. For this reason, half of the `detectable' lensed TDEs may be missed from the survey for the double image cases due to the time-sequential effect. 

\section{Results}
\label{sec:results}

Figure \ref{fig:rate} shows the expected detection rate of the lensed TDEs ($\dot{N}_{\rm TDE}^{\mathcal{l}}$) by surveys with different limiting magnitudes $m_{\rm lim}$ in the u-band (blue), g-band (red), and z-band (black), respectively. The solid and dashed lines represent the results obtained for the cases with $\beta_{\rm d}=1.00$ and $\beta_{\rm d}=1.85$, respectively. Obviously the larger the limiting magnitude is, the larger the amount of lensed TDEs that can be detected. We also note here that in the detection of lensed TDEs with relatively shorter wavelength are more effective due to the spectrum of the mock events. 

The expected lensed TDE detection rate $\dot{N}_{\rm TDE}^{\mathcal{l}}$ can exceed $1$\,yr$^{-1}$ in the case with $\beta_{\rm d}=1.00$ with the limiting magnitude of $m_{\rm lim}\gtrsim 22.9$, $23.1$, and $23.7 $\,mag in the u-, g-, and r-band, respectively. If alternatively set $\beta_{\rm d}=1.85$, then the detection of lensed TDEs can be much larger (e.g., $\sim 31.9$, $30.9$, and $22.7 $ for $m_{\rm lim} \sim 23.5$ in  the u-, g-, and z-band, respectively), because the predicted peak TDE luminosities are substantially larger than those with $\beta_{\rm d} =1$ (see Fig.~\ref{fig:LC}). In the case with $\beta_{\rm d}=1.85$, one expects to detect at least one lensed TDE per year by an all sky survey with $m_{\rm lim}=21.3$, $21.2$, and $21.5$\,mag in the u-, g-, z-band, respectively. Moreover, with the strong power of future transient sky surveys, for example $m_{\rm lim}\gtrsim 25-26 $ in the u-, g-, z-bands, such as Rubin \citep[e.g.,][]{2020ApJ...890...73B,2023ApJS..268...13B}, the total detection rate may be upto several tens to hundreds per year.

Figure~\ref{fig:td} shows the time-delay $t_{\rm d}$ distribution between the first and second images of the strongly gravitationally lensed TDEs. As seen from this figure, the median value of $t_{\rm d}$ is $\sim 16$ days 
and 
roughly 70\% of the lensed TDEs have time delays exceeding 10 days, a duration typically associated with the time interval between the commencement of observations ($t_{\rm obs}$) and the occurrence of the peaks ($t_{\rm peak}$) in the LCs of TDEs. Such large intrinsic time-delays enable the early-warning for one to capture the complete LCs in the optical bands and even to perform pre-peak UV/optical/X-ray spectroscopy of the lensed TDEs. The supplementary observations of subsequent lensed images may contribute invaluable insights to various aspects of the TDE research, which will be further elaborated in the discussion section.

The detectable lensed TDEs may have different properties from those TDEs in the parent sample. For illustration, Figure~\ref{fig:para} shows the mass and redshift distribution of those MBHs in the detectable lensed TDEs (left panel) and the radius and mass of the disrupted stars (right panel) by an all-sky survey with the limiting magnitude of $\sim 23.5$ mag in the g-band (the required limiting magnitude for detecting one lensed TDE per year even if $\beta_{\rm d}=1$). (The distributions are similar in other bands and in the case of $\beta_{\rm d}=1.85$.) As seen from this figure, the median redshift and mass of the MBHs in the detectable lensed TDEs are $z_{\rm s}\sim 1.73 $ and $2\times 10^{5} M_\odot$, respectively. 

The masses of the corresponding disrupted stars in the lensed TDEs peak at $m_\star \sim 1.20 M_\odot$ (with the median of $1.26 M_\odot$), which is relatively large compared with the 
input median mass of disrupted stars without consideration of their life time. The reason is that the lensed TDEs are mostly located at relatively high redshift ($z_{\rm s}\sim 1-2$) and only those bright ones (with large-mass disrupted star) can be detected due to the magnitude limitation. Lensed TDEs with small-mass disrupted stars are relatively fainter and thus are less likely to be detected. Here we also noticed that compared with the mock mass distribution of disrupted stars in this paper, the available measurements of the disrupted star masses for the observed TDEs may vary by a factor of $\sim 0.1-10$ or more, due to  differet settings of the LC models. For example, as for the TDE source PS1-10jh, the star mass estimated by different LC fitting codes, i.e., MOSFiT, TDEmass, and TiDE 
\citep[e.g.,][]{2019ApJ...872..151M,2020ApJ...904...73R,2023PASP..135c4102K}, can be very different, which is $\sim 0.41/1.80/27.12 M_{\odot}$ respectively. More detailed comparison on these three codes can be seen in \citet{2023PASP..135j4102K}.

\section{Conclusions and Discussions}
\label{sec:conclusion}

In this paper, we investigate the detectability of strongly gravitationally lensed TDEs and estimate their detection rate by transient surveys, by adopting a simple wind+disk model to predict TDE LCs and the SIE+external shear model for lenses to estimate the optical depth for lensing. We find that the requisit limiting magnitude for an all-sky transient survey to detect at least one resolvable lensed TDE are $\sim 22.9$, $23.1$, and $23.7$\,mag (or $\sim 21.3$, $21.2$, and $21.5$\,mag) in the u-, g-, and z-band, respectively, if $\beta_{\rm d}=1.00$ (or $1.85$). If the limiting magnitude of the survey can reach $\sim 25-26$\,mag in these bands, the detection rate of the lensed TDEs can be upto several tens to hundreds per year.

We demonstrate that the LCs of multiple images in a single band of a lensed TDE are intrinsically the same except they are magnified by the lensing galaxy with constant factors. The combination of the LCs from different images may lead to the enhancement of the signal-to-noise ratio compared with those unlensed TDEs at the same redshift and therefore help us to understand the origin of the optical radiation in such a remote and  complex system. Moreover, if the time-delay $t_{\rm d}$ between two images is larger than the time between the observation starting-time $t_{\rm obs}$ and the peak time $t_{\rm peak}$ of the first image, then the signal of the first image can be taken as an early-warning of the second image. For example, if one could observe the full LC of the second image, one may estimate $t_{\rm min}$ by extrapolating the LC and therefore constrain the physical processes in the early stage of TDEs. Moreover, this early warning may enable 
pre-peak UV/optical/X-ray spectroscopy observation, which is however rare in literature. These additional information may be useful in several aspects. First, the emission line spectra in the UV/optical band are detectable, for example, the Balmer and/or Bowen lines, facilitates the immediate classification of TDEs based on their spectral line characteristics \citep[e.g.,][]{2012Natur.485..217G,2021ApJ...908....4V,2023ApJ...952L..35Z}. Secondly, if the pre-peak evolution of emission lines are observable, it becomes feasible to determine the changes in velocity and opacity of the super-Eddington outflow by examining linewidth and intensity, respectively. This information is crucial for comprehending the underlying accretion physics in TDEs. Thirdly, an early pre-peak X-ray radiation is predicted in the reprocessing scenario \citep[e.g.,][]{2009MNRAS.400.2070S, 2011MNRAS.410..359L, 2018ApJ...859L..20D,2022MNRAS.516.2833B}, while no such signal will be produced in the shock+circularization \citep[e.g.,][]{1990ApJ...351...38C, 2013MNRAS.435.1809S, 2022MNRAS.510.1627A} scenario. {Therefore, by ulitilizing the current and future powerful X-ray detectors, such as Chandra \citep{2002PASP..114....1W}, Swift X-Ray Telescope (Swift/XRT) \citep{2005SSRv..120..165B}, Neutron Star Interior Composition Explorer Mission (NICER) \citep{2016SPIE.9905E..1HG}, XMM-Newton \citep{2001A&A...365L...1J} and Einstein Probe \citep{2015arXiv150607735Y}, the early-warning of X-ray allows the pre-peak observation of the second lensed TDE image and therefore provide clues to the identification of origin of flares in TDE.}

In addition, we estimate the parameter space for the MBHs and the disrupted stars of those `detectable' lensed TDEs by an all-sky survey. We find that the detectable lensed TDEs may have different properties from those TDEs in the total sample. For the very first ones discoved by sky survey(s) with a limiting magnitude of $\sim 23.5$ or less, the disrupted stars may have a mass around $1.26M_\odot$.

Note that we only consider simple cases for demonstration in this paper, however, there should be more complexities in the real observations of such special transient events. For example, we only consider the galaxy as the lens object and adopt the simple SIE plus external shear lens model but ignore the cluster-lensing. The inclusion of cluster-lensing may enhance the detection rate of lensed TDEs as the magnification factors by cluster-lensing are larger, leading to brighter lensed TDEs. Moreover, the intrinsic LCs of TDEs adopted in this paper are obtained from an over-simplified phemenonlogical model and the luminosity of TDEs can vary significantly with different choice of the model parameters, for example, $\beta_{\rm d}$, which may directly change our estimation on the detection rate by orders of magnitude. 

We further note that the detection rate of lensed TDEs by surveys in the near future may be substantially smaller than the estimate above because of the following factors. First, the detection rate of lensed TDEs is dependent on the sky-coverage and the cadence of the sky survey(s) to search for them. In this paper, for simplicity, we consider the ideal all-sky survey, with single-epoch cadence of $\lesssim 3-4$ days to reach a limiting magnitude of $\gtrsim 21$\,mag or $25-26$\,mag (in the u-, g-, and z-bands), that can at least detect one lensed TDE per year or several tens to hundreds lensed TDEs. {For instance, Rubin can approach such a cadence with a limiting magnitude of $\sim 25$\,mag but it is not an all-sky survey (with the sky-coverage of $\sim 20000$\,deg$^2$). Therefore, Rubin can detect about half of the predicted number presented in Section~\ref{sec:results}, which is $\sim 5$ or $60$ lensed TDEs for $\beta_{\rm d}=1.0$ or $1.85$ cases, respectively.} For other telescopes, for example, the wide field survey telescope (WFST) \citep{2023SCPMA..6609512W} and Pan-STARRS need to coadd several exposures of single-visits to reach $m_{\rm lim}\sim 25$ \citep{2016arXiv161205560C}, 
therefore the requisite total time for a $\sim 20000$\,$\rm deg^2$ transient survey with such limiting magnitude is about $\sim 100-200$ days. Then in total the detection rate of lensed TDEs may be smaller than that estimated in this paper by a factor of $\sim 2-5$.

Besides, as for the strategy part, we have assumed that all the lensed host galaxies of TDEs can be identified by space-borne galaxy surveys, i.e., $f_{\rm host}=1$. However, we find that for lensed TDEs $f_{\rm host}\sim 0.5$, which is very similar to the AGN/MBH channel case discussed in our previous work on the identifiable lensed hosts of lensed GW sources \citep[e.g.,][]{2022ApJ...940...17C,2023MNRAS.524.2954M,2023MNRAS.518.6183M}. Therefore, the predicted lensed TDE detection rate in our paper may decrease by a factor of $\sim 2$. However, we note here that $f_{\rm host} $ is mostly limited by the human-inspection criteria (see \citet{2022ApJ...940...17C} for detailed description ) on the shape recognition of the lensed galaxies rather than the limiting magnitude. With the rapid development of the strong machine-learning technique on the galaxy-galaxy lensing identification \citep[e.g.,][]{2017Natur.548..555H, 2018MNRAS.473.3895L, 2023ApJ...954...68Z}, one may identify the lensed hosts more accurately, making $f_{\rm host}\sim 1$ as expected. Therefore, we are optimicstic of the detection of lensed TDEs and {hopefully}  
may help to resolve several important physical aspects involved in TDEs.

\section*{acknowledgement}
We are grateful to Qingjuan Yu (KIAA at PKU) for her contribution to this work. We also thank Ning Jiang (USTC), Chichuan Jin (NAOC), and Erlin Qiao (NAOC) for their suggestions to improve this work. This work is partly supported by the National Natural Science Foundation of China (grant nos. 12273050, 11991052), the Strategic Priority Research Program of the Chinese Academy of Sciences (grant No. XDB0550300), and the National Key Program for Science and Technology Research and Development (grant nos. 2020YFC2201400 and 2022YFC2205201).

\bibliographystyle{aasjournal}
\bibliography{ref.bib}

\end{document}